\documentclass[aps,prx,superscriptaddress,floatfix,amsmath,amsfonts,twocolumn]{revtex4-2}  
\usepackage[russian,english]{babel}

\usepackage{enumitem}
\usepackage{amsmath}
\usepackage{graphicx}
\usepackage{epsfig}
\usepackage{bm}
\usepackage{bbold}
\usepackage{amssymb}
\usepackage{color}
\usepackage[normalem]{ulem}  %

\begin{document}

\title{Persistent polarization oscillations in ring-shape polariton condensates} 

\author{A.V. Yulin}
\affiliation{Department of Physics, ITMO University, Saint Petersburg 197101, Russia}

\author{E. S. Sedov}
\affiliation{Russian Quantum Center, Skolkovo, Moscow 143025, Russia}
\affiliation{Spin-Optics laboratory, St. Petersburg State University, St. Petersburg 198504, Russia}
\affiliation{Vladimir State University, Vladimir 600000, Russia}

\author{A.V. Kavokin}
\affiliation{Westlake University, School of Science, 18 Shilongshan Road, Hangzhou 310024, Zhejiang Province, China}
\affiliation{Westlake Institute for Advanced Study, Institute of Natural Sciences, 18 Shilongshan Road, Hangzhou 310024, Zhejiang Province, China}
\affiliation{Spin-Optics laboratory, St. Petersburg State University, St. Petersburg 198504, Russia}

\author{I.A. Shelykh}
\affiliation{Science Institute, University of Iceland, Dunhagi 3, IS-107, Reykjavik, Iceland}
\affiliation{Abrikosov Center for Theoretical Physics, MIPT, Dolgoprudnyi, Moscow Region 141701, Russia}
\affiliation{Department of Physics, ITMO University, Saint Petersburg 197101, Russia}

\date{\today}

\begin{abstract}
We predict the limit cycle solution for a ring-shape bosonic condensate of exciton-polaritons confined in an optically induced rotating trap. The limit cycle manifests itself with polarization oscillations on a characteristic timescale of tens of picoseconds. The effect arises due to the interplay between orbital motion and the polarization degree of freedom. It is specific to spinor bosonic condensates and would be absent in a scalar case, where a bi-stability of stationary solutions would be observed instead. This work offers a tool of initialisation and control of qubits based on superpositions of polariton condensates characterised by different topologic charges.
\end{abstract}

\maketitle

\section{Introduction}

Strongly coupled exciton-photon systems are known for their pronounced optical nonlinearities, enabling the dynamic control of light. A typical representative of this class of systems is an optical microcavity with one or several semiconductor quantum wells characterised with resonant excitonic transitions~\cite{KavokinMicrocavities}. In such a structure, if the energies of confined photons and excitons are tuned close to resonance, their interaction leads to the formation of half-light, half-matter hybrid modes referred to as exciton polaritons. The field of polaritonics keeps attracting the enhanced attention of the research community, as it offers a convenient testing platform for various quantum coherent and nonlinear effects~\cite{Carusotto2013}. A remarkable phenomenon that came to the focus of attention in the new century is the formation of dynamical Bose-Einstein condensates (BECs) of polaritons at exceptionally high temperatures~\cite{kasprzak2006bose,balili2007bose}. Similar to regular BECs formed at the thermodynamic equilibrium, polariton condensates clearly demonstrate  such properties as macroscopic coherence and superfluidity, arising from the repulsive polariton-polariton interactions~\cite{carusotto2004probing,wouters2008excitations, amo2009superfluidity,wouters2010superfluidity,lerario2017room}. The effects of polariton condensation can be employed for the creation of a new class of nanoscale coherent light sources known as polariton lasers~\cite{pau1996observation,christopoulos2007room,schneider2013electrically}.

The properties of the dynamic polariton condensates in confined geometries are governed by the intrinsic spatio-temporal dynamics of polaritons, that may include formation of persistent polariton currents corresponding to quantized vortex states ~\cite{kavokin2003polariton,carusotto2004probing,cristofolini2013optical,chestnov2016nonlinear, lerario2017room,Lukoshkin2018,Sedov2021,SciRep1122382,Lukoshkin2023,sanvitto2010persistent}. An intriguing parallel can be drawn between superfluid polariton vortices and Abrikosov vortices that strongly affect properties of superconductors of the second type \cite{abrikosov1957magnetic}. Similar analogies extend to nonlinear optics~\cite{snyder1992stable} and atomic Bose-Einstein condensates~\cite{Pitaevsky1961,kivshar1998dark}.  

Recently, it has been suggested that a superposition of polariton vortices with opposite topological charges can serve as a robust and scalable  qubit~\cite{Xue2021,Kavokin2022,barrat2023superfluid,ricco2023qubit}. This makes especially important creation of a theoretical framework for the description of the properties of polariton vortex states, including their nonlinear dynamics~\cite{lagoudakis2008quantized,lagoudakis2009observation,sanvitto2010persistent,nardin2011hydrodynamic,lagoudakis2011probing}.
Central to this problem is the development of protocols allowing for the control of vortex topological charges~\cite{assmann2012all,dall2014creation,nardin2010selective, manni2011spin,yulin2016spontaneous,kwon2019direct,yulin2020spinning}

\begin{figure}[tbh!]
\centering\includegraphics[width=\columnwidth]{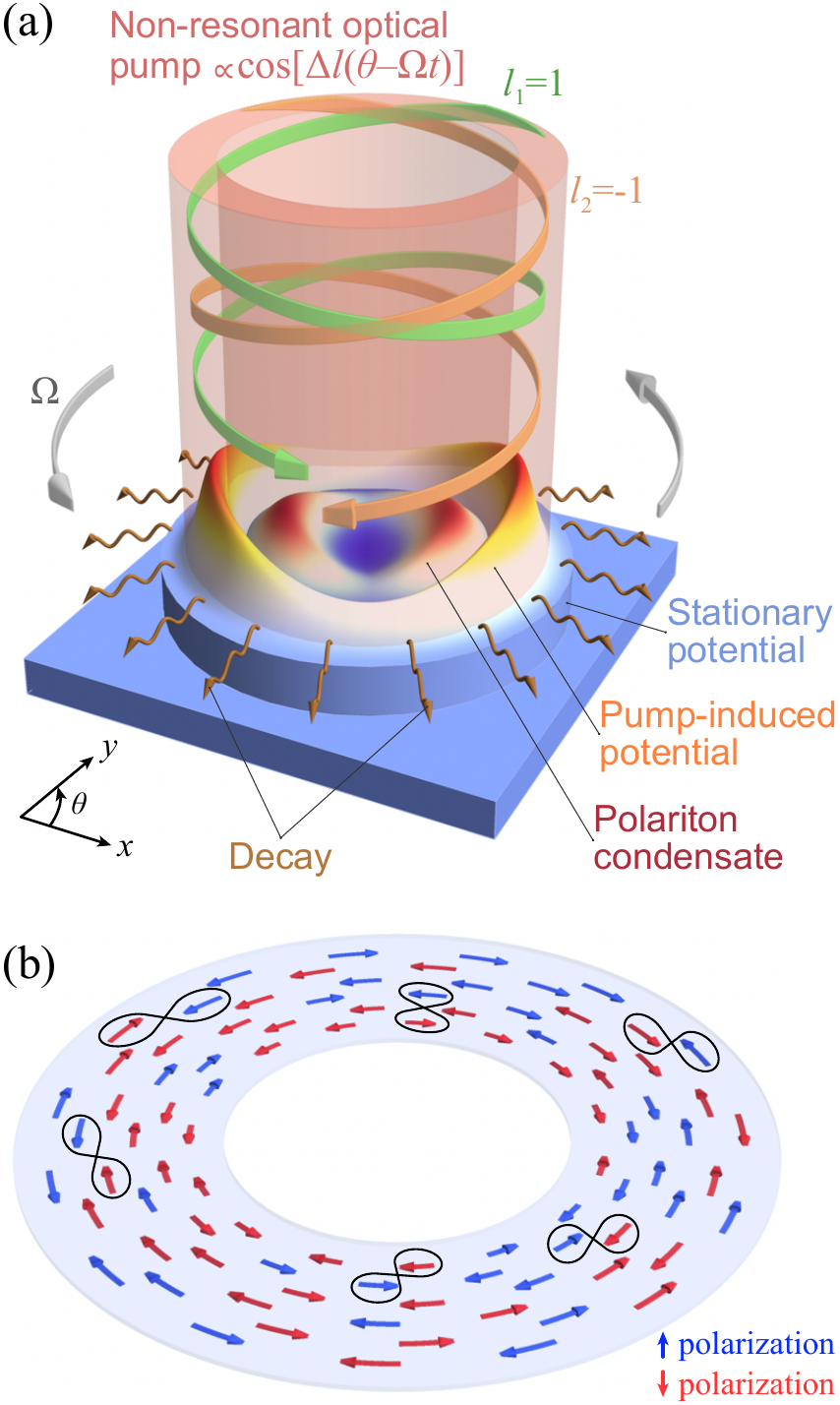}
\caption{
(a) Schematic of the excitation of a polariton condensate in a rotating optical potential.
The potential is created by two optical beams with angular momenta $l_1 = 1$ and $l_2 = -1$, characterized by different frequencies, $\Omega = \omega _1 - \omega _2$.
The polariton condensate exists under the balance of the optical pump and decay. 
(b) Schematic of the coupling of counter-propagating polariton flows via polarization due to the TE-TM splitting.
$\uparrow$~polarized polaritons (blue arrows) propagating in the counterclockwise direction couple to $\downarrow$~polarized polaritons (red arrows) propagating in the clockwise direction. 
\label{FIG_Scheme}}
\end{figure} 

It should be noticed that while quantized vortices can exist in spatially uniform planar microcavities, the introduction of a confining potential can stabilize vortices and offer an additional tool to control their properties~\cite{PhysRevB74155311,1010634983832,NatPhys6860,LSA879,PhysRevB91195308, PhysRevB97235303,PhysRevB101245309,PhysRevB107045302,PhysRevB91195308,PhysRevB97235303, PhysRevB101245309,PhysRevB107045302}. In particular, introduction of chirality into the confining potential facilitates the excitation of polariton vortex states with a precise control over their topological charge~\cite{Dall2014,Lukoshkin2018ACSPhot,Lukoshkin2018,Sedov2021}. The problem of spinning of a polariton condensate by optically induced rotating potentials has been recently studied theoretically and experimentally ~\cite{lagoudakis2022,Fraser,yulin2023spin,yulin2023vorticity}. It has been demonstrated that in the case of a single component (scalar) condensate, in the vicinity of the condensation threshold, the formation of a condensate is governed exclusively by the properties of the fastest growing linear mode.  ``The winner takes all'' scenario~\cite{yulin2023vorticity,Faltskog1980}, where the fastest growing mode is the first to reach the nonlinear stage at which it depletes the pump and suppresses the growth of the other modes, is realized in this regime. 

In the present work, we analyze the mechanisms of mode selection in spinor condensates, where the polarization degree of freedom is accounted for.
Polaritons of different polarizations exhibit distinct behaviors under the influence of spin-orbit interaction effects, which link the polarization of polaritons to their ballistic propagation within the microcavity plane. Among such effects are the optical spin Hall effect~\cite{PhysRevLett95136601,NatPhys3628}, zitterbewegung of polaritons~\cite{PhysRevB97245312,LSA122047}, topological spin Meissner effect~\cite{PhysRevB94115407} and etc. Of particular interest are the effects of spin-orbit interaction in structures with annular geometry~\cite{PhysRevLett102046407,PhysRevB94115407,PhysRevB106245309}. Spin-orbit interaction occurs under conditions of splitting of polarization states of polaritons, which can be induced by both the inherent optical anisotropy of the structure~\cite{,PhysRevB545852,PhysRevB64045312} and external influences~\cite{PhysRevLett106257401,PhysRevB81125311,CommPhys2165}. In layered structures of optical microcavities, the most significant source of spin-orbit interaction is the splitting of transverse electric (TE) and transverse magnetic (TM) polariton modes~\cite{PhysRevB595082}.

In this paper we show that in certain regimes, ``the winner takes all'' scenario can be realized in the spinor condensate as well. 
However, we also show that in the specific range of the rotation speeds, the superposition of two modes characterized by different frequencies and polarizations becomes stable. This gives rise to the limit cycle regime characterized by periodic polarization oscillations. Quite remarkably, these oscillations are not accompanied by the oscillations of the total polariton occupancy. This is demonstrated by direct numerical simulations of the two-dimensional (2D) generalized Gross-Pitaevskii equation describing the dynamics of a driven-dissipative polariton condensate. Additionally, we develop a simple perturbation theory based on the coupled modes approach~\cite{Sedov2021,SciRep1122382}, which provides a qualitative description of the predicted phenomena.     

\section{The formalism}

We consider the system consisting of an axially symmetric semiconductor microcavity that is incoherently pumped by two linearly polarized Laguerre-Gaussian optical beams. These beams possess different angular momenta, $l_1$ and $l_2$, $\Delta l=l_1-l_2\neq 1$, and are slightly detuned in frequency, so that $\Omega=\omega _1\neq\omega _2$. The resulting spatial profile of the pump can be represented as
\begin{equation}
 P (t,r, \theta) = P_{s}(r) + P_{r}(r) \cos\Delta l (\theta +\Omega t),   
\end{equation}
where $r$ and $\theta$ denote the radial and angular coordinates, and $P_s(r)$ and $P_r(r)$ describe the radial distributions of the non-rotating and rotating components of the pump.
{In further consideration, we threat the rotating pump excited by a superposition of two optical Laguerre-Gaussian beams having the angular momenta ${l_{1} = 1}$ and ${l_{2} = -1}$, thus $\Delta l = 2$.}

The optical pump excites the reservoir of incoherent excitons.
Due to the repulsive polariton-polariton interactions, the excitonic reservoir acts as an effective rotating complex trapping potential for the polariton condensate, with its imaginary part describing the stimulated relaxation of the excitons into the polariton state. 
We also account for an additional real rotationally symmetric trapping potential stemming from the patterning  of a microcavity and stabilizing the polariton confinement. A schematic plot of the system is given in Fig.~\ref{FIG_Scheme}(a).  

As typically the internal reservoir dynamics is much slower then the relaxation from it into the condensate, the reservoir can be adiabatically eliminated \cite{Keeling2008} and the system is described by the two coupled generalized Gross-Pitaevskii equations for order parameters $\Psi_{\uparrow, \downarrow}$ corresponding to two circular polarizations:
\begin{widetext}
\begin{multline}
i \hbar \partial_{\bf{t}} \Psi_{\uparrow, \downarrow} = \frac{\hbar^2}{2m_{\text{eff}}} (\partial_{\bf{x}}^2+\partial_{\bf{y}}^2) \Psi_{\uparrow, \downarrow} -\left( {\bf{V}} +i\hbar \gamma \right) \Psi_{\uparrow, \downarrow}+\frac{\hbar^2  \beta }{2m_{\text{eff}}}(\partial_{\bf{x}} \pm i\partial_{\bf{y}})^2 \Psi_{\downarrow, \uparrow}  \\
-\hbar (H|\Psi_{\uparrow, \downarrow}|^2 +\tilde H |\Psi_{\downarrow, \uparrow}|^2)\Psi_{\uparrow, \downarrow}
- \left( \frac{\hbar (G_2-i G_1) P}{\Gamma+2 G_1 |\Psi_{\uparrow, \downarrow}|^2} +  \frac{\hbar \tilde G_2 P}{\Gamma+2 G_1 |\Psi_{\downarrow, \uparrow}|^2} \right) \Psi_{\uparrow, \downarrow}. \label{master_eq} 
\end{multline}
\end{widetext}
The first term on the right-hand side describes the dispersion of polaritons characterized by effective mass~$m_{\text{eff}}$. The second term accounts for the presence of an external conservative potential $\bf{V}$, created through microstruturing, together with a finite lifetime of polaritons characterized by an energy broadening ~$\gamma$. The third term corresponds to the TE-TM splitting, with normalized splitting magnitude $\beta$ being proportional to the difference of the effective masses of TE and TM polarized polaritons \cite{Flayac2010}. The fourth term describes the blueshift of a polariton condensate due to repulsive polariton-polariton interactions, with coefficient $H$ describing the interaction between polaritons with same circular polarizations, $\tilde{H}$ the interaction between polaritons with opposite circular polarizations. 

The last term describes polariton-reservoir coupling, with its real part corresponding to the reservoir-induced polariton blueshift, and its imaginary part corresponding to the condensate gain stemming from the stimulated relaxation from the reservoir to the condensate. The exciton reservoir is driven by the linear polarized optical pump of intensity~$P$, which excites equally both circular polarization components. 
The parameter $G_1$ is the scattering rate from the reservoir to the condensate, and $\Gamma$ is the relaxation rate of the reservoir excitons. The coefficient $G_2$ defines the reservoir-induced blue shift of the condensate. 
The small red shift caused by the interaction of polaritons with excitons of opposite polarization is characterized by the coefficient $\tilde G_2$.

For further analysis, we introduce dimensionless variables, scaling time in units of the inverse polariton dissipation rate, $t_0=\gamma^{-1}$, $t\rightarrow t/t_0$, the spatial coordinates in units of  $l_0 = \sqrt{\left. \hbar \right/ 2 \gamma m_{\text{eff}}}$, $\mathbf{r}\rightarrow\mathbf{r}/l_0$, and the order parameter $\Psi_{\uparrow, \downarrow}\rightarrow\sqrt{\left. 2G_1\right/ \Gamma} \psi_{\uparrow, \downarrow}$. Then Eq.~\eqref{master_eq} can be written as following:
\begin{widetext}
\begin{multline}
i \partial_{t} \psi_{\uparrow, \downarrow} = \frac{1}{2}(\partial_{x}^2+\partial_{y}^2) \psi_{\uparrow, \downarrow} -\left( \frac{i}{2}+V\right) \psi_{\uparrow, \downarrow}+\beta (\partial_{x} \pm i\partial_{y})^2 \psi_{\downarrow, \uparrow}   \\
- (h|\psi_{\uparrow, \downarrow}|^2 +\tilde h |\psi_{\downarrow, \uparrow}|^2)\psi_{\uparrow, \downarrow}
- \left( \frac{(1-i \epsilon) p}{1+|\psi_{\uparrow, \downarrow}|^2} +  \frac{ g p}{1+|\psi_{\downarrow, \uparrow}|^2} \right) \psi_{\uparrow, \downarrow}, \label{master_eq_normalized} 
\end{multline}
\end{widetext}
where $V = {\bf{V}} / \hbar \gamma$ is the normalized stationary potential, 
${\epsilon=G_1 / G_2}$ is the ration of the effective gain to the frequency shift caused by the reservoir, 
${g=\tilde G_2 / G_2}$ is the ration of the frequency shifts due to interactions of polaritons with the reservoir excitons of the opposite and the same polarizations, 
${p = P G_2 / \gamma \Gamma}$ is the normalized incoherent pump, ${h = H \Gamma / 2 \gamma G_1}$ and ${\tilde h = \tilde H \Gamma / 2 \gamma G_1}$ are the relative strengths of the self- and cross-polarization polariton-polariton interactions. 

In our simulations, we use the following values of the parameters, typical for polariton systems: ${\beta = 0.05}$, ${\epsilon = 0.33}$, and ${g=-0.1}$. We focus on the scenario in which the nonlinearity stemming from the reservoir is dominating and take the coefficients  $h=0.018$ and $\tilde h=-0.001$. {The corresponding values of dimensional parameters in Eq.~\eqref{master_eq} are given in~\cite{RealParamValues}.} We also noted that in the vicinity of the condensation threshold, variations of the values of these parameters did not qualitatively change the picture.
External confining potential was taken in the form: $V=V_0 \left\{ \exp[-(r-R_V)^8/W_V^8]\right. + \left. \exp[-(r+R_V)^8/W_V^8] \right\}$ with $V_0=7$, $R_V=2.25$ and $W_V=1$.
{This corresponds to a ring-shape confinement with the height of the potential, radius, and width of the ring being about $0.28 \,\text{meV}$, ${9.89 \, \mu \text{m}}$, and  ${4.39 \, \mu \text{m}}$, respectively.}

As discussed above, the pump is the combination of the static symmetric term $p_s(r)$ and the term $p_r(r) \cos[2(\theta -\Omega t)]$ rotating with angular velocity~$\Omega$. 
The radial distribution of the static pump is taken in the following form: $p_s (r) = p_{s0} \left\{ \exp[-(r-R_s)^2/W_s^2] \right. + \left. \exp[-(r+R_s)^2 / W_s^2] \right\}$. For our simulations, we take $p_{s0}=3.45$, $R_s=0.9$ and $W_s=0.25$. {This corresponds to a ring-shape pump of intensity $0.17 \,\text{ps}^{-1} \, \mu\text{m}^{-2}$ with radius and width of the ring being of about  ${3.96 \, \mu \text{m}}$ and  ${1.1 \, \mu \text{m}}$, respectively.} In this case, the confinement of polaritons results from the combination of the external conservative potential~$V$ and static  potential induced by the pump. Selecting a smaller radius of the pump annulus compared to the radius of the external potential enhances the efficiency of the incoherent excitation of polaritons. Such a combination ensures that only condensates with angular indices of either $+1$ or $-1$ are formed in the system. 

The radial distribution of the rotating pump component $p_r(r)$ is taken to be equal to that of the stationary one $p_s(r)=p_r(r)$. In our simulations, we take the following values of the parameters of the rotating pump: $p_{r0}=0.3$, $R_r=0.9$, and $W_r=0.25$. {This corresponds to a ring-shape pump of intensity $0.015 \,\text{ps}^{-1} \, \mu\text{m}^{-2}$ with radius and width of the ring being of about  ${3.96 \, \mu \text{m}}$, and ${1.1 \, \mu \text{m}}$, respectively. }

\section{Results of 2D modelling}

\subsection{Stationary regime}

\begin{figure}[tb!]
\begin{center}
\includegraphics[width=0.9\columnwidth]{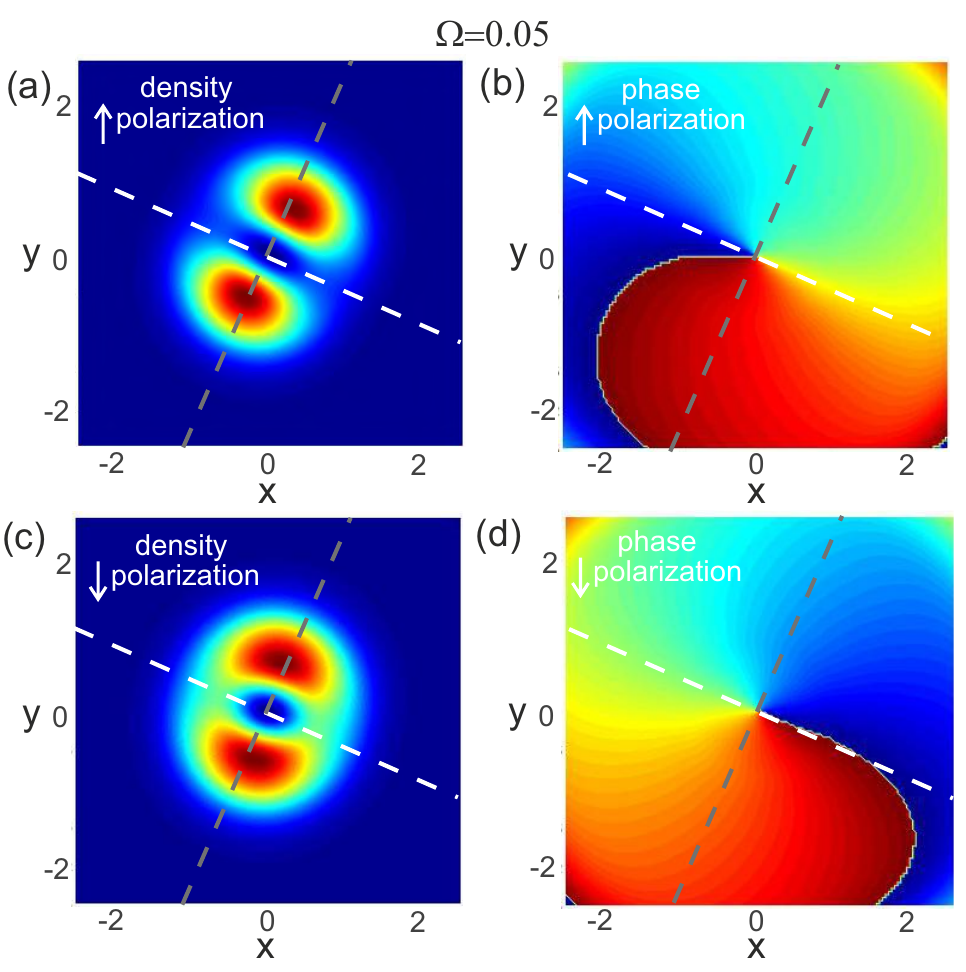}
\end{center}
\caption{The snapshots showing instantaneous profiles of the intensity and the phase distributions of the polariton field in opposite circular polarizations: $\uparrow$ for  (a),(b) and $\downarrow$ for (c),(d).
The angular velocity of the potential is~$\Omega=0.05$. The dashed lines show the symmetry axes, with the white line passing through the minima and the gray line through the maxima of the rotating potential. 
\label{FIG_01_IntPhase}}
\end{figure}

We performed numerical studies of the dynamics of polariton states forming in the considered geometry in the presence of a complex rotating potential. We covered a broad range of the rotation velocities, wherein we observed the formation of the stationary states stably developing from a weak noise. 
Our results reveal that the density distributions of the polaritons in clockwise and counterclockwise circular polarizations ($\uparrow$ and $\downarrow$) depend on the angular velocity of the potential rotation.   
For instance, at a relatively small rotation velocity $\Omega=0.05$, we observed the distinct lobes in the density distribution of $\uparrow$~polarization (as depicted in Fig.~\ref{FIG_01_IntPhase}(a)), indicating that in this polarization, the condensate is formed by two counterpropagating waves of comparable amplitudes. The phase distribution of the order parameter, illustrated in Fig.~\ref{FIG_01_IntPhase}(b), clearly demonstrates that the winding number for this polarization of the condensate is equal~to~$1$. {Let us note here that in the case of two counter-propagating waves, the topological charge is determined by the wave with the larger amplitude.}

In the opposite polarization, $\downarrow$, the density distribution similarly exhibits two lobes, albeit notably less pronounced. This discrepancy indicates that in this polarization, the waves with angular momenta $m=\pm 1$ forming the state possess rather different amplitudes. It's important to highlight that the phase gradient is directed oppositely in the opposite circular polarization, meaning that polaritons of the $\uparrow$ and $\downarrow$ polarization flow in opposite directions. Furthermore, it's worth noting that the lobes in both polarizations are oriented along the symmetry axis, passing through the maxima of the rotating potential. 

\begin{figure}[tb!]
\begin{center}
\includegraphics[width=0.9\columnwidth]{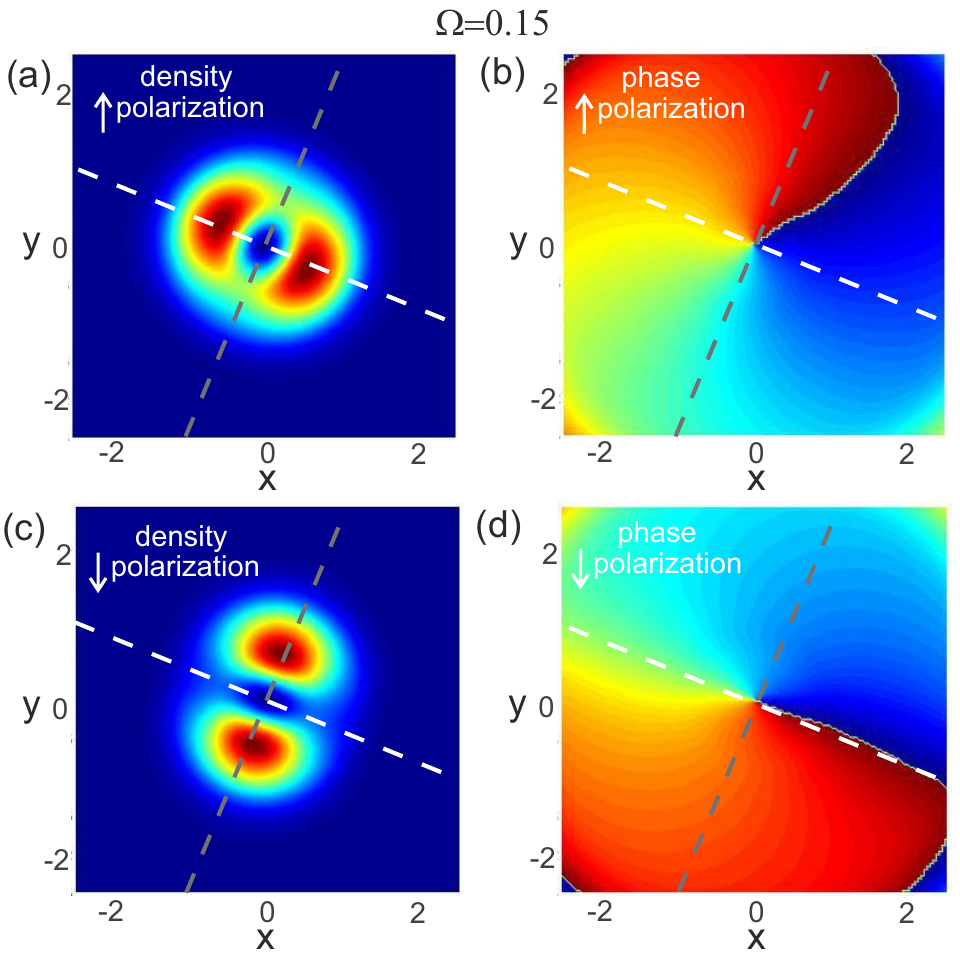}
\end{center}
\caption{
The snapshots showing instantaneous profiles of the intensity and the phase distributions of the polariton field in opposite circular polarizations: $\uparrow$ for  (a),(b) and $\downarrow$ for (c),(d).
The angular velocity of the potential is~$\Omega=0.15$.
The dashed lines show the symmetry axes, with the white line passing through the minima and the gray line through the maxima of the rotating potential. \label{FIG_03_IntPhase}}
\end{figure}

For  higher rotation velocities, the distributions of the polarization intensities are different, as can be seen in Fig.~\ref{FIG_03_IntPhase}, which shows the snapshots of the stationary polariton densities and phases for~$\Omega=0.15$. 
Here the lobes in the $\uparrow$~polarization are less pronounced than in the $\downarrow$~polarization. Notably, the orientation of the lobes  differs as well. Namely, the lobes in the $\uparrow$~polarization  now align along the symmetry axis that traverses the minima of the potential.

\begin{figure}[tb!]
\centering\includegraphics[width=\columnwidth]{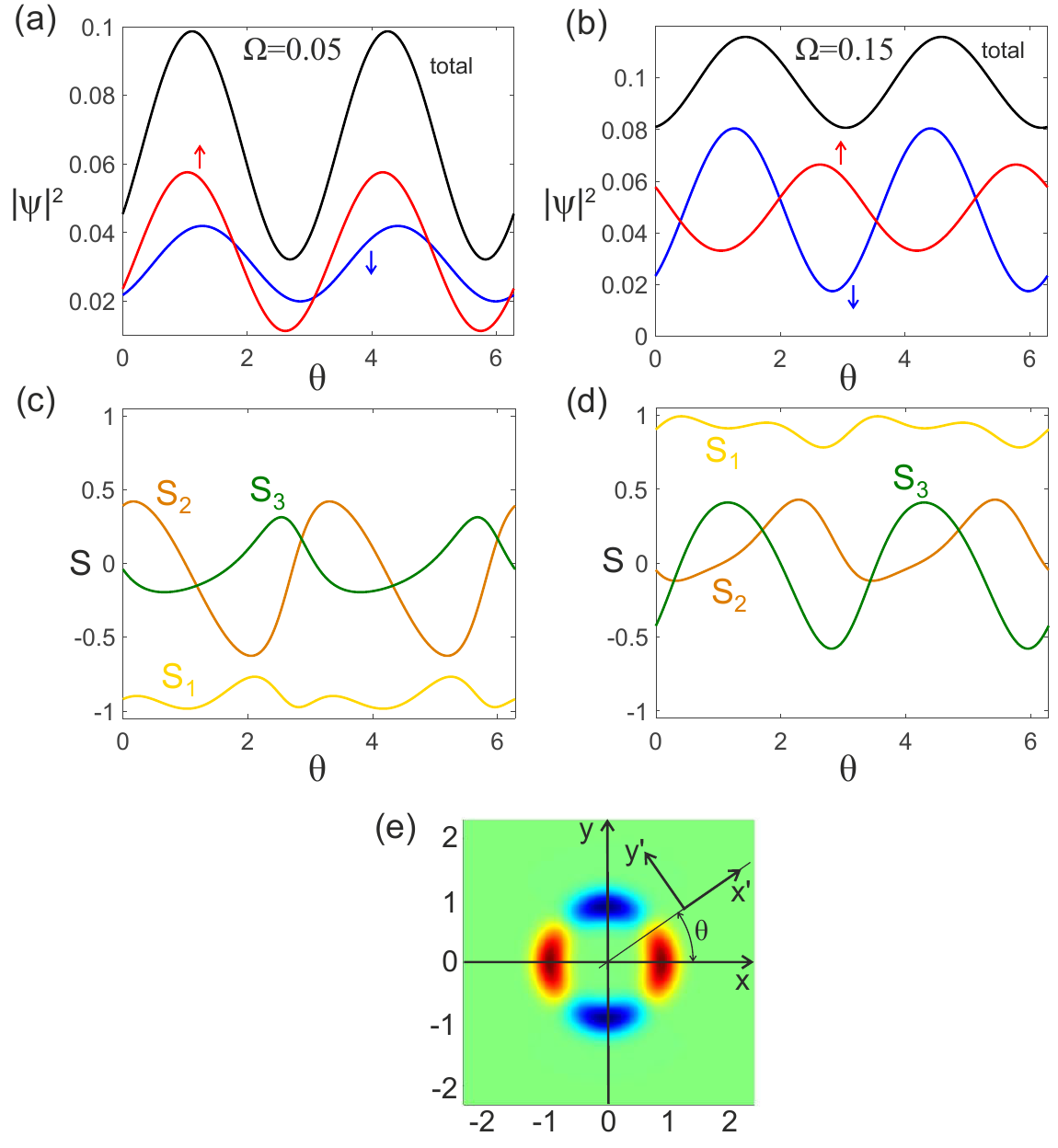}
\caption{(a) The angular distribution of the total density of the stationary state of the polariton condensate $|\psi|^2$ (black curves) as well as the condensate polarization components $|\psi _{\uparrow}|^2$ (red curves) and $|\psi _{\downarrow}|^2$ (blue curves) for the rotation angular velocities $\Omega=0.05$ (a) and $\Omega=0.15$ (b).
(c), (d) The angular dependencies of the normalized Stokes parameters for the corresponding angular velocities. The dependencies are plotted for $r=0.8$. 
The Stokes parameters are calculated in the rotating reference frame $x^{\prime} o y^{\prime}$, see panel~(e). 
(e) The spatial distribution of the rotating potential. $x o y$~is the laboratory reference frame, wile $x^{\prime} o y^{\prime}$~is the rotating reference frame.  
\label{FIG_Stokes_stationary}}
\end{figure}

For discussing the polarization properties of polaritons, it is convenient to introduce the normalized three-component Stokes vector~$\vec{S} = (S_1,S_2,S_3)$, which components are determined as follows: ${S_1 = 2 \text{Re} (\psi ^* _{\uparrow} \psi _{\downarrow}) / |\psi|^2}$, ${S_2 = -2\text{Im} (\psi ^* _{\uparrow} \psi _{\downarrow}) / |\psi|^2}$ and ${S_3 = (|\psi _{\uparrow}|^2 -  |\psi _{\downarrow}|^2) / |\psi|^2}$,  where ${|\psi|^2 = |\psi _{\uparrow}|^2 + |\psi _{\downarrow}|^2}$ is total condensate density.
It's essential to note that we calculate the Stokes components in the rotating frame~$x'oy'$, see Fig.~\ref{FIG_Stokes_stationary}(e).
In this frame, the $S_1$ component corresponds to the $x'$ / $y'$ linear polarizations, while in the laboratory frame $xoy$, it characterizes the tangential and radial polarization components of the condensate.
The other components, $S_2$ and $S_3$, are responsible for diagonal/antidiagonal and clockwise/counterclockwise circular polarizations, respectively.

The azimuthal distribution of the densities, as well as the Stokes vector components of the polariton condensates in the stationary state, are depicted in Fig.~\ref{FIG_Stokes_stationary} for the discussed rotation velocities of the potential~$\Omega$.
It is evident that in the regime corresponding to the lower rotation velocity ${\Omega = 0.05}$ (panels (a) and (c)), the polaiton field is predominantly radially polarized, with $S_1$ close to~$-1$.
Conversely, in the regime corresponding to the higher rotation velocity ${\Omega = 0.15}$ (panels (b) and (d)), the polaiton field is primarily tangentially polarized, with $S_1$ close to~$+1$.

We also calculated the normalized angular momenta in the clockwise and counterclockwise ($\uparrow$ and $\downarrow$) polarizations defined as follows: ${ {\cal M}_{\uparrow, \downarrow} =  \left. M_{\uparrow, \downarrow} \right/ N_{\uparrow, \downarrow}}$, where $M_{\uparrow, \downarrow}= \text{Im} \int \psi_{\uparrow, \downarrow}^{*} (\partial_x+i\partial_y)\psi_{\uparrow, \downarrow} dx dy$ is the actual angular momentum, while  $N_{\uparrow, \downarrow} = \int |\psi_{\uparrow, \downarrow}|^2 dxdy$ stands for the number of polaritons in the corresponding polarization. The total normalized  angular momentum can be calculated as follows:  ${ {\cal M}_{\text{tot}}=\left. (M_{\uparrow}+M_{\downarrow}) \right/ N_{\text{tot}} }$, where ${N_{\text{tot}} = N_{\uparrow} + N_{\downarrow}}$ is the total number of polaritons in the condensate. The dependencies of the normalized angular momenta of the stationary condensates in the rotating potentials as functions of the rotation velocity~$\Omega$ are shown in Fig.~\ref{FIG_AM_ndiff_dpendns}(a).

\subsection{Breathing regime}

\begin{figure}[tb!]
\centering\includegraphics[width=\columnwidth]{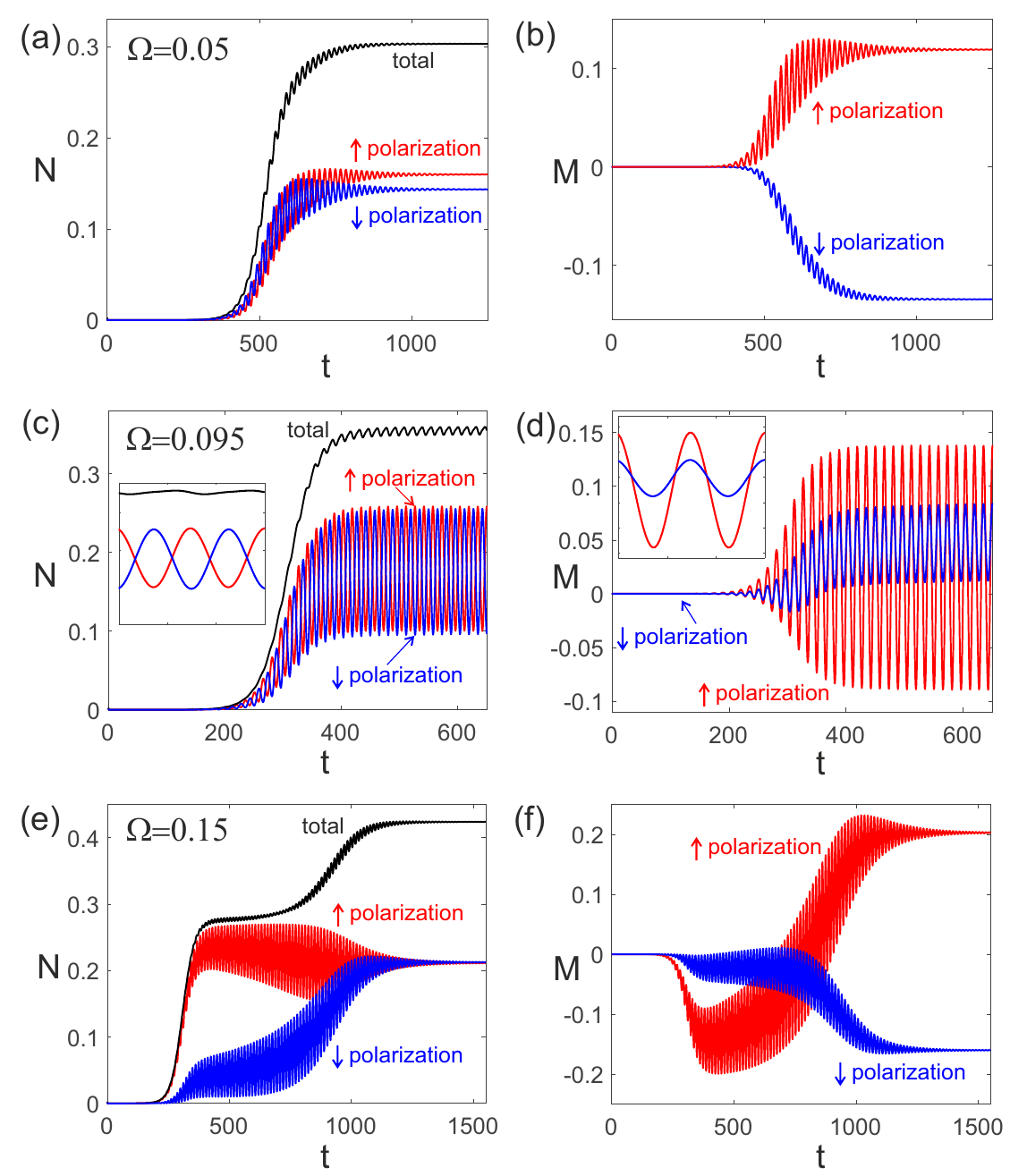}
\caption{ The temporal dependencies of the polariton numbers $N$ and the angular momenta $M$ are shown in panels (a), (c), (e) and (b), (d), (f) correspondingly. The potential rotation angular velocity is $\Omega=0.05$ for (a), (b); $\Omega=0.095$ for (c), (d) and $\Omega=0.15$ for (e), (f). 
\label{FIG_dynamics}}
\end{figure}

An important observation emerges: within a specific range of the angular velocities of the trapping potential $\Omega$, the system supports only breathing steady state solutions. The typical dynamics of polaritons is illustrated in Fig.~\ref{FIG_dynamics}, where the evolution of the number of polaritons and the angular momentum of the polariton states in different polarizations is presented. As shown in panel (a), for relatively slow rotation, the number of particles grows in both polarizations until the steady state is reached. A similar behavior is observed for angular momenta, see panel~(b).  

\begin{figure}[tb!]
\centering\includegraphics[width=0.9\columnwidth]{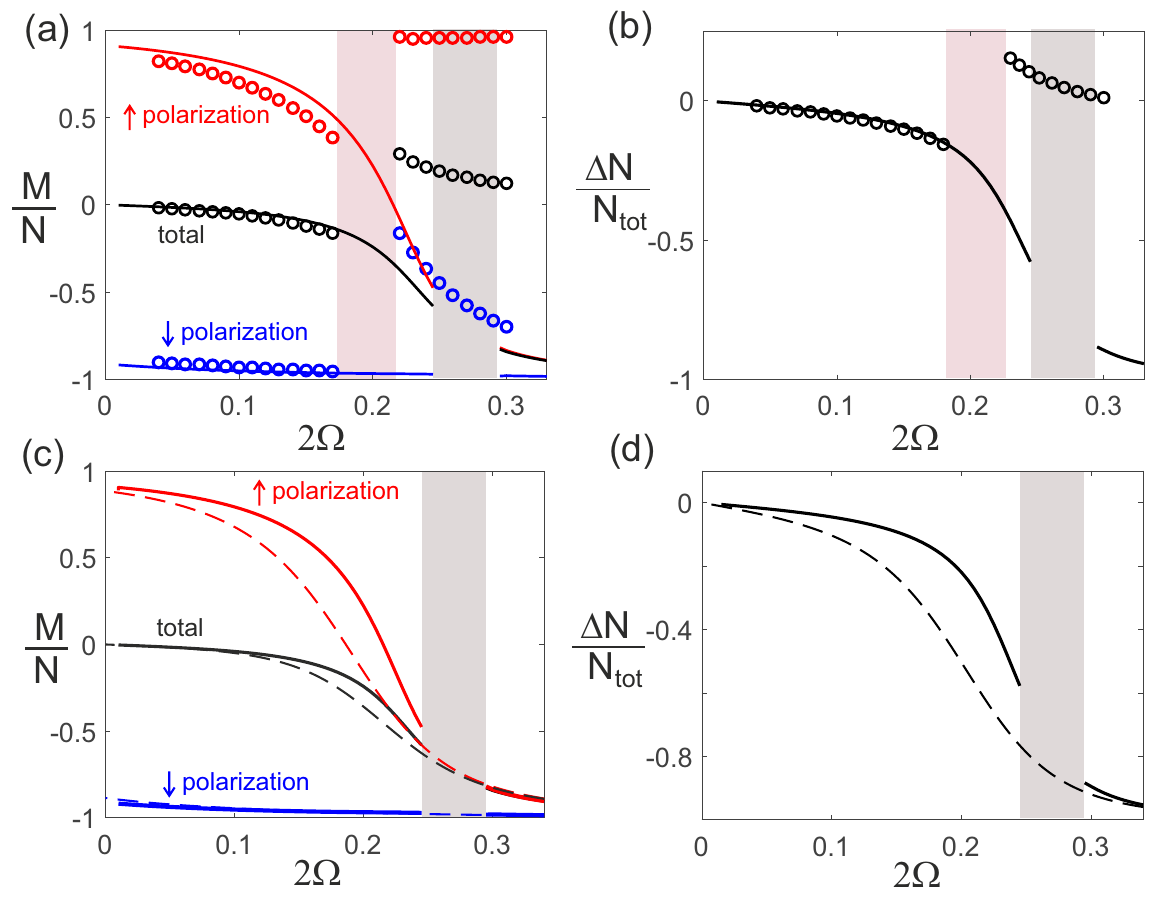}
\caption{The dependencies of the normalized angular momenta $\mathcal{M_{\uparrow, \downarrow, \text{tot}}}$ (a),(c) and the normalized polarization occupation difference $\Delta N /N_{\text{tot}} = (N_{\uparrow} - N_{\downarrow})/(N_{\uparrow} + N_{\downarrow})$ (b),(d) on the potential angular velocity~$\Omega$. The open circles in (a) and (b) show the quantities obtained from 2D numerical simulations. The solid lines in all panels correspond to numerical simulations based on nonlinear equations from the coupled mode theory. The dashed lines in (c) and (d) are for the fastest growing linear mode calculated from the coupled mode theory. Pink and gray rectangles indicate the ranges of the rotation velocity~$\Omega$, predicted from 2D numerical simulations (pink) and from the coupled mode approach (gray), where the steady state exhibits oscillatory behavior.
\label{FIG_AM_ndiff_dpendns}}
\end{figure}

However, within the angular velocity range indicated by the pink rectangle in Figs.~\ref{FIG_AM_ndiff_dpendns}(a,b), the dynamics takes on a markedly different character. 
This becomes evident in Fig.~\ref{FIG_dynamics}(c), where the steady state manifests oscillations in the number of polaritons in both polarizations. A spatial counterpart to this periodic inter-polarization transfer of polaritons has been recently observed in polariton waveguides~\cite{rozas2021effects}.

The angular momenta in the $\uparrow$ and $\downarrow$~polarizations oscillate in phase, as depicted in panel (d), thus yielding pronounced oscillations in the total angular momentum of polaritons. Intriguingly, the numbers of polaritons in $\uparrow$ and $\downarrow$~polarization oscillate in anti-phase, leading to only minor oscillations in the total number of polaritons. 

This phenomenon arises when the polarization oscillation period $T_{\text{osc}}$ is much shorter compared to the characteristic timescale of evolution of the polariton density~$T_{\text{dns}}$, ${T_{\text{osc}} < T_{\text{dns}}}$. The latter is determined by how much the pump intensity exceeds the condensation threshold intensity.
In particular, for intensities that are sufficiently close to the threshold, this condition is always satisfied. The total number of the polaritons thus cannot follow the rapid periodic transfer of the polaritons between the polarizations, and, therefore, the total number of the polaritons does exhibit pronounced oscillations. Some estimates are given in Appendix~\ref{oscillating behavior} to explain the discussed effect. 

\begin{figure}[tb!]
\centering\includegraphics[width=\columnwidth]{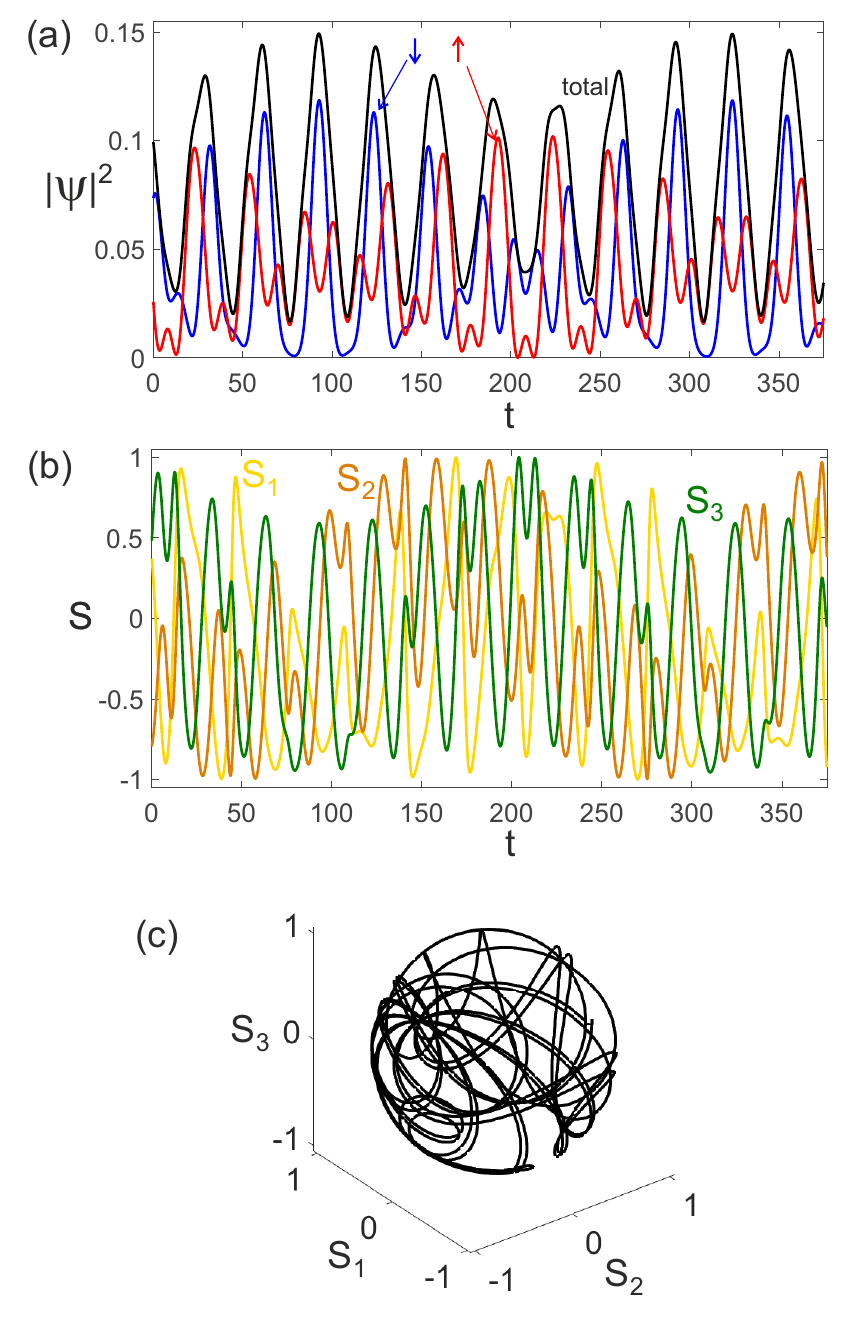}
\caption{Evolution in time of the polariton densities (a) and the Stokes parameters (b) in the oscillatory regime for the polariton field at $x=0.8$, $y=0$. (c)~The trajectory of the normalized Stokes vector on the Poincar\'{e} sphere.
\label{FIG_polariza_oscill}}
\end{figure}

In the oscillatory regime, the temporal dynamics of the Stokes vector becomes quasi-periodic, see Fig.~\ref{FIG_polariza_oscill}.
The oscillations of the polarization measured at a fixed point in real space (specifically, at $x=0.8$, $y=0$) are accompanied by oscillations of intensity.
The trajectory of the Stokes vector on the Poincar\'{e} sphere demonstrates intricate behavior and covers the entire sphere's surface, indicating that during the evolution, the polariton condensate undergoes oscillations between linear and circular polarization states.

With an increase in the rotation velocity, suppression of the oscillations becomes evident. As seen in Fig.~\ref{FIG_dynamics}(e,f), after some transitional processes, a stationary state (shown in Fig.~\ref{FIG_03_IntPhase}) eventually emerges. Nevertheless, it's worth noting that the transitional stage is notably more intricate compared to that observed at lower angular velocities (sf.~panels (a,b) and (e,f)). 

In the next section, we develop a perturbation theory that allows us to analyze this phenomenon.

\section{Coupled modes approximation}

In order to comprehend the effects reported in the previous section, we developed a straightforward perturbation theory based on the observation that only the modes possessing angular indices of~$\pm 1$ can be effectively excited in the system. We also restrict our consideration to the case where the spatial structure of the modes is predominantly defined by the real parts of stationary potentials, supposing that  dissipative and nonlinear effects, the rotating potential and the TE-TM splitting can be treated as small perturbations. This allows us to represent the field as a sum of the four modes, each characterized by its polarization and angular momentum. The total field can be represented as follows:
\begin{multline}
\left( \begin{array}{cc}
     \psi_{\uparrow} (r, \theta, t) \\
     \psi_{\downarrow} (r, \theta, t) 
\end{array} \right) \\
= \rho(r) e^{i\omega_0 t} 
\left( \begin{array}{cc}
     A_{\uparrow +}(t)e^{i\theta} + A_{\uparrow -}(t)e^{-i\theta}  \\
     A_{\downarrow +}(t) E^{i\theta} + A_{\downarrow -}(t) e^{-i\theta}  
\end{array} \right), \label{anzats} 
\end{multline}
where $\omega_0$ is the eigenfrequency of the unperturbed mode, and $\rho(r)$ describes their radial structure of the mode. We normalize the latter as ${4\pi \int r\rho(r)dr=1}$ ensuring that the total number of polaritons is equal to $N_{\text{tot}}=|A_{\uparrow +}|^2+|A_{\uparrow -}|^2+|A_{\downarrow +}|^2+|A_{\downarrow -}|^2$.

To eliminate explicit time dependencies of coefficients, we introduce new variables 
%
\begin{subequations}
\label{EqNewVars}
\begin{equation*}
C_{\uparrow +} = A_{\uparrow +}, \quad
C_{\uparrow -} = A_{\uparrow -}e^{2i\Omega t}, \eqno{(\ref{EqNewVars}\text{a,b})}
\end{equation*}
\begin{equation*}
C_{\downarrow +} = A_{\downarrow +}e^{-2i\Omega t}, \quad
C_{\downarrow -} = A_{\downarrow }. \eqno{(\ref{EqNewVars}\text{c,d})}
\end{equation*}
\end{subequations}
The dynamic equations can be then written as:
\begin{widetext}
\begin{subequations}
\label{EqCoupled_modeC}
\begin{eqnarray}
\dot C_{\uparrow +} = &-& \left[ \gamma_{0}  +(\alpha_d-i\alpha)\left(|C_{\uparrow +}|^2+2|C_{\uparrow -}|^2\right)\right]C_{\uparrow +} + \nonumber \\
&+&i\alpha_x \left[ \left( |C_{\downarrow +}|^2+|C_{\downarrow -}|^2 \right)C_{\uparrow +} + C_{\uparrow -}C_{\downarrow +}C_{\downarrow -}^{*} \right] 
+ i\eta C_{\uparrow -} + i\sigma C_{\downarrow -}
\label{coupled_mode_C1} \\ %
\dot C_{\uparrow -} = &-& \left[\gamma_{0} -2i\Omega +(\alpha_d-i\alpha)\left(|C_{\uparrow -}|^2+2|C_{\uparrow +}|^2\right)\right] C_{\uparrow -} + \nonumber \\
&+& i\alpha_x\left[ \left( |C_{\downarrow +}|^2+|C_{\downarrow -}|^2 \right)C_{\uparrow -} + C_{\uparrow +}C_{\downarrow -}C_{\downarrow +}^{*} \right] 
+i\eta  C_{\uparrow +} 
\label{coupled_mode_C2} \\ %
\dot C_{\downarrow +} = &-& \left[ \gamma_{0} + 2i\Omega +(\alpha_d-i\alpha)\left(|C_{\downarrow +}|^2+2|C_{\downarrow -}|^2\right)\right] C_{\downarrow +}+ \nonumber \\
&+& i\alpha_x\left[ \left( |C_{\uparrow +}|^2+|C_{\uparrow -}|^2 \right)C_{\downarrow +} + C_{\downarrow -}C_{\uparrow +}C_{\uparrow -}^{*} \right]
+i\eta C_{\downarrow -} \label{coupled_mode_C3} \\ %
\dot C_{\downarrow -} = &-& \left[ \gamma_{0}  +(\alpha_d-i\alpha)\left(|C_{\downarrow -}|^2+2|C_{\downarrow +}|^2\right)\right] C_{\downarrow -}+ \nonumber \\
&+&i\alpha_x \left[ \left( |C_{\uparrow +}|^2+|C_{\uparrow -}|^2 \right)C_{\downarrow -} + C_{\downarrow +}C_{\uparrow -}C_{\uparrow +}^{*} \right] 
+i\eta  C_{\downarrow +} + i\sigma C_{\uparrow +}.
\label{coupled_mode_C4}
\end{eqnarray}
\end{subequations}
\end{widetext}
The details of derivation of Eqs.~\eqref{EqCoupled_modeC} are given in~Appendix~\ref{AppCoupledModeEqs}.

The effective losses of the modes $\gamma_0$, the coupling strength between the modes of the same polarization but opposite angular momenta $\eta$, the effective TE-TM splitting $\sigma$, the dissipative nonlinearity $\alpha_d$, the conservative cubic nonlinearity $\alpha$ as well as the conservative cross-polarization nonlinearity $\alpha_x$ coefficients can be derived by fitting the special series of 2D simulations. This procedure gives the following estimates for the coefficient values:
$\gamma_0=0.006$, $\sigma=0.21$, $\alpha_d=0.081$, $\alpha=0.281$, $\alpha_x=-0.027$. The coefficient characterizing the rotating potential is expressed as $\eta=\eta_r+i \eta_i$, with $\eta_r=0.052$, $\eta_i=-0.017$.

One should pay special attention to the last terms in the right-hand side of Eqs.~\eqref{coupled_mode_C1} and~\eqref{coupled_mode_C4}. 
These terms elucidate the role of the TE-TM splitting in the polariton evolution in the ring geometry, which consists in the coupling of counter-propagating polariton flows via the polarization degree of freedom~\cite{yulin2023spin,SciRep1122382}.
Namely, it couples counterclockwise propagating ($+$) flows of $\uparrow$~polarized polaritons with clockwise propagating~($-$) flows of $\downarrow$~polarized polaritons (see Fig.~\ref{FIG_Scheme}(b)).

We can now define the angular momenta and the occupancies of the polariton state in $\uparrow$ and $\downarrow$ polarizations via the amplitudes of the polariton modes as follows: 
\begin{subequations}
\label{EqAngMandNviaAmplC}
\begin{eqnarray}
&&M_{\uparrow, \downarrow} = |C_{\uparrow, \downarrow \, +}|^2 - |C_{\uparrow, \downarrow \, -}|^2, \\
&&N_{\uparrow, \downarrow}= |C_{\uparrow, \downarrow \, +}|^2 + |C_{\uparrow, \downarrow \, -}|^2.
\end{eqnarray}
\end{subequations}
With the notation~\eqref{EqAngMandNviaAmplC}, the quantities $\mathcal{M}_{\text{tot}}$ and $N_{\text{tot}}$ defined in the previous section keep their meaning.

The results of the numerical solution of Eqs.~\eqref{EqCoupled_modeC} are shown in Fig.~\ref{FIG_AM_ndiff_dpendns} by solid lines.
The dependencies of the normalized angular momenta $\mathcal{M} _{\uparrow, \downarrow, \text{tot}}$ and the normalized population difference of the orthogonal circularly polarized modes ${\Delta N/N_{\text{tot}}}$, where $\Delta N = N_{\uparrow}-N_{\downarrow}$, on the angular velocity $\Omega$ of the rotating potential are shown in Figs.~\ref{FIG_AM_ndiff_dpendns}(a) and~\ref{FIG_AM_ndiff_dpendns}(a), respectively.
The quantity ${\Delta N/N_{\text{tot}}}$ actually represents the degree of circular polarization of the polariton condensate.

It is evident that the results of the coupled mode approximation align well with the results of the full-scale 2D simulations. We have also observed that the fitting accuracy tends to improve at lower pump intensities, particularly when the system is just above the condensation threshold. It's worth noting  that as the system approaches the condensation threshold, the time required for the formation of the stationary state becomes considerably elongated, rendering the simulations more time-consuming.

It is worth mentioning that the transition to the oscillatory regime is also qualitatively reproduced by the coupled mode approximation. However, the range of the angular velocities where the oscillatory regime occurs (gray rectangles in Figs.~\ref{FIG_AM_ndiff_dpendns}(a,b)), does not coincide with the one obtained from 2D simulations. This discrepancy could be attributed to the sensitivity of the dynamics of the system to the values of the parameters as well as to the neglect of certain effects in the coupled mode approach, such as the contributions of non-resonant modes and the frequency dependence of effective losses.

Another noteworthy aspect is that at low angular velocities of the rotating potential, the stationary states obtained from both 2D simulations and the coupled mode approach exhibit the same structural characteristics. However, at higher angular velocities, this congruence no longer holds, and the coupled mode approach predicts a distinct structure of the polariton condensates. A possible explanation is that the stationary state anticipated by the coupled mode approach exists in the 2D model, but it proves to be unstable with quite a relatively small increment. This instability might account for the scenario depicted in Fig.~\ref{FIG_dynamics}(e,f), where an oscillatory state initially forms with the averages quite close to that predicted by the coupled mode approach for the chosen parameters. Howeverm in the coupled mode approach, we observe the decay of the oscillations, whereas in full-scale 2D simulations, transition to a different state takes place. It is acknowledged that further research is required to comprehensively understand this intricate dynamic behavior, which is left for future exploration. 

{Based on our numerical simulations, we propose that in the vicinity of the condensation threshold, the formation of the stationary state follows the ``winner takes all'' scenario. When the initial conditions involve a randomly distributed low-intensity field, the mode with the fastest growth rate is the first to reach the nonlinear stage. However, both the nonlinear depletion of the pump and the linear increment are small, and consequently, the nonlinearity does affect the structure of the modes. It is worth noting that the rotating pump couples the modes of the unperturbed problem, resulting in the eigenmodes becoming compositions of the eigenmodes of the axially symmetric problem. When we refer to the ``structure of the eigenmode'', we are describing the relative amplitudes of the waves that compose the eigenmode.

In the scenario under consideration, the fastest growing mode tends to dominate and suppress the other modes before its amplitude reaches the stationary value. Consequently, the structure of the nonlinear stationary state gets inherited from the fastest growing linear mode. This scenario holds well in the scalar case. For certain values of the rotation velocities, this concept also applies to the vector case, as detailed in Appendix~\ref{Lin_Dyn}, where we analyze the linear eigenmodes. 
Further insights into the perturbative analysis of the nonlinear stage of the dynamics can be found in Appendix~\ref{Nonlinear_perturbation}, where we derive an approximate expression for the amplitude of the stationary state.

In the vector case, at specific rotation velocities~$\Omega$, the fastest growing mode exhibits a structure where a majority of polaritons are in the $\uparrow$ (or, for a different sign of $\Omega$, in $\downarrow$) polarization. Consequently, the mode with dominant $\uparrow$ ($\downarrow$) polarization cannot effectively suppress the mode where a significant portion of polaritons are in the $\downarrow$ ($\uparrow$) polarization. This phenomenon is illustrated in Fig.~\ref{FIG_AM_ndiff_dpendns}(b), where the polariton portion in one polarization decreases before transitioning to the oscillation regime.
To substantiate this observation, an analysis of the stability of the non-oscillating states is required. This analysis is presented in Appendix~\ref{oscillating behavior}. Additionally, in the same Appendix~\ref{oscillating behavior}, it is demonstrated that the oscillating states can be seen as the coexistence of two sub-states of different frequencies. The structures of the sub-state are similar to the structures of the fastest and second fastest growing modes. }

\section{Conclusion}

We have studied the polarization dynamics of polariton condensates in the presence of a rotating potential. Our main attention was focused on low density condensates that are formed when the pump intensity is close to the condensation threshold.
We have shown that during the initial linear stage of the condensation, the rotating potential leads to the formation of super-modes, resulting from the hybridization of modes affected by TE-TM splitting with those unaffected by it.
The most pronounced hybridization occurs at the resonant angular velocity of the rotating potential, which matches the strength of TE-TM splitting. Due to the dissipative component of the rotating potential, the imaginary parts of these super-modes differ. Interestingly, the mode with the highest frequency also exhibits the highest growth rate.

We have demonstrated that in a non-conservative system, the ``winner takes all'' scenario can be realized. In this case, during the nonlinear stage of condensation, the fastest growing mode suppresses the other modes. Consequently, the stationary state is predominantly determined by the structure of the fastest growing linear mode, provided that the pump intensity is close to the condensation threshold and the density of the polariton state does not influence the field distributions.

The main result of this study is the finding of an original dynamics of the system characterised by persistent oscillations of the polarization. This limt cycle regime can be achieved even for the pump intensity slightly above the threshold. It occurs for certain angular velocities where the system is in a state where the fastest growing mode predominantly consists of polaritons of a specific circular polarization. Consequently, this mode cannot effectively deplete the gain created by the pump in the opposite polarization. As a result, the remaining pump is sufficient to support the growth of another super-mode. Thus, the final state comprises two coexisting super-modes with different polarizations.

The frequency difference between these modes leads to oscillations in the number of polaritons in clockwise and counterclockwise circular polarizations. The total angular momentum of polaritons undergoes periodic oscillations as well, but the total number of polaritons remains close to constant over time.

The results of direct numerical solutions of the two-dimensional coupled generalized Gross-Pitaevskii equations are supported by a simple semi-analytical theory based on coupled mode approximation.

\begin{acknowledgments}
{I.A.S. acknowledges financial support from Icelandic Research Fund (Rannis, the project ``Hybrid polaritonics''), ``Priority 2030 Academic Leadership Program'' and ``Goszadanie no. 2019-1246''. I.A.S. also acknowledges support from Ministry of Science and Higher Education of Russian Federation (Agreement No. 075-15-2021-589). The numerical simulations performed by A.V.Y. were financially supported by the Russian Science Foundation (project No. 23-72-00031).
The work of E.S.S. was carried out within the state assignment in the field of scientific activity of the RF Ministry of Science and Higher Education (theme FZUN-2020-0013, state assignment of VlSU). 
A.V.K. and E.S.S. acknowledge Saint-Petersburg State University for
the financial support (research grant No.~94030557).}
\end{acknowledgments}

\appendix

\section{Derivation of equations in the coupled modes approximation \label{AppCoupledModeEqs}}

Let us develop a simplified model for revealing the evolution of the polariton condensate in a rotating potential.
We start with the dimensionless Gross-Pitaevskii equation~\eqref{master_eq_normalized}.
Given the assumption that the nonlinear effects in the considered system are weak (which is valid at the pump power close to the condensation threshold), we limit ourselves to considering nonlinearity up to the third order in amplitude.
We expand the nonlinear terms in Taylor series up to the cubic terms in~$\psi$ and obtain
\begin{widetext}
\begin{multline}
\left[ \frac{(1-i \epsilon) p}{1+|\psi_{\uparrow, \downarrow}|^2} +  \frac{ g p}{1+|\psi_{\downarrow, \uparrow}|^2} \right] \psi_{\uparrow, \downarrow} 
+(h|\psi_{\uparrow, \downarrow}|^2 +\tilde h |\psi_{\downarrow, \uparrow}|^2)\psi_{\uparrow, \downarrow}  \\
\approx (1+g-i \epsilon)p\psi_{\uparrow, \downarrow}+ (h-p+i \epsilon p)|\psi_{\uparrow, \downarrow}|^2\psi_{\uparrow, \downarrow} +(\tilde h- gp)|\psi_{\downarrow, \uparrow}|^2\psi_{\uparrow, \downarrow}.
\label{NL_expans} 
\end{multline}
\end{widetext}
It is worth mentioning that the effective nonlinearity comprises the contribution from both polariton-polariton interactions and the depletion of the incoherent pump. Typically, these contributions are of different signs, leading to a variation in the sign of the effective nonlinearity based on which contribution dominates. In our findings, the qualitative outcomes remain consistent regardless of the sign of the nonlinear frequency shift. 
For our current investigation, we focus on the scenario where the primary nonlinearity arises from the depletion of the reservoir, resulting in a nonlinear red shift experienced by polaritons.

To describe the dynamics within this approximation, we need to derive ordinary differential equations for the amplitudes $A_{\uparrow,\downarrow \pm}$.
We substitute the decomposition~\eqref{anzats} into the master equation~\eqref{master_eq_normalized} and, using the approximation~\eqref{NL_expans}, obtain the equations for the mode amplitudes~$A_{\uparrow \downarrow, \pm}$:
\begin{widetext}
\begin{subequations}
\label{EqCoupled_modeA}
\begin{eqnarray}
\dot A_{\uparrow +} = &-&\left(\gamma_{0} +(\alpha_d-i\alpha)\left(|A_{\uparrow +}|^2+2|A_{\uparrow -}|^2\right)\right)A_{\uparrow +} \nonumber \\
&+&i\alpha_x\left( \left( |A_{\downarrow +}|^2+|A_{\downarrow -}|^2 \right)A_{\uparrow +} + A_{\uparrow -}A_{\downarrow +}A_{\downarrow -}^{*} \right)+i\eta e^{2i\Omega t} A_{\uparrow -} + i\sigma A_{\downarrow -} ,
 \label{coupled_mode_A1} \\ %
\dot A_{\uparrow -} = &-& \left(\gamma_{0} +(\alpha_d-i\alpha)\left(|A_{\uparrow -}|^2+2|A_{\uparrow +}|^2\right)\right)A_{\uparrow -} \nonumber \\ 
&+& i\alpha_x\left( \left( |A_{\downarrow +}|^2+|A_{\downarrow -}|^2 \right)A_{\uparrow -} + A_{\uparrow +}A_{\downarrow -}A_{\downarrow +}^{*} \right)+i\eta e^{-2i\Omega t} A_{\uparrow +} ,
\label{coupled_mode_A2} \\ %
\dot A_{\downarrow +} = &-& \left(\gamma_{0} +(\alpha_d-i\alpha)\left(|A_{\downarrow +}|^2+2|A_{\downarrow -}|^2\right)\right)A_{\downarrow +} \nonumber \\
&+& i\alpha_x\left( \left( |A_{\uparrow +}|^2+|A_{\uparrow -}|^2 \right)A_{\downarrow +} + A_{\downarrow -}A_{\uparrow +}A_{\uparrow -}^{*} \right)+i\eta e^{2i\Omega t} A_{\downarrow -} ,
 \label{coupled_mode_A3} \\ %
\dot A_{\downarrow -} = &-&\left(\gamma_{0} +(\alpha_d-i\alpha)\left(|A_{\downarrow -}|^2+2|A_{\downarrow +}|^2\right)\right)A_{\downarrow -}  \nonumber \\
&+& i\alpha_x\left( \left( |A_{\uparrow +}|^2+|A_{\uparrow -}|^2 \right)A_{\downarrow -} + A_{\downarrow +}A_{\uparrow -}A_{\uparrow +}^{*} \right)+i\eta e^{-2i\Omega t} A_{\downarrow +} + i\sigma A_{\uparrow +}.
 \label{coupled_mode_A4}
\end{eqnarray}
\end{subequations}
\end{widetext}
After changing the variables in~\eqref{EqCoupled_modeA} according to~\eqref{EqNewVars}, we arrive at Eq.~\eqref{EqCoupled_modeC}.

\section{Dynamics of linear excitations \label{Lin_Dyn}}

To comprehend the behavior of polaritons, we conduct a linear analysis using the coupled modes approximation. This involves neglecting the nonlinear terms in Eqs.~\eqref{EqCoupled_modeC} and subsequently writing the equation for the vector $\vec C=\left( C_{\uparrow +}, C_{\uparrow -}, C_{\downarrow +}, C_{\downarrow +}\right)^{\text{T}}$ of the mode amplitudes in a matrix form:
\begin{eqnarray}
\label{EqEigEq}
\frac{d}{dt} \vec C = i \hat L \vec C, \label{Vec_lin}
\end{eqnarray}
where 
\begin{eqnarray}
\hat L=
\left( \begin{array}{cccc}
  i \gamma_{0}&  \eta                &  0                   &   \sigma  \\
  \eta      &  i \gamma_{0}+2\Omega &  0                   &    0  \\
  0          &  0                    & i \gamma_{0} -2\Omega &    \eta \\
  \sigma    &  0                    & \eta                &    i\gamma_{0}
\end{array} \right). \nonumber
\end{eqnarray}
The eigenvalues of $\hat L$ correspond to the eigenfrequencies of the polariton modes, with the imaginary part indicating the decay rate of the mode. While exact solutions of Eq.~\eqref{EqEigEq} can be obtained analytically, they are cumbersome and are not included in the text of this paper. 

\begin{figure}[tb!]
\centering\includegraphics[width=\columnwidth]{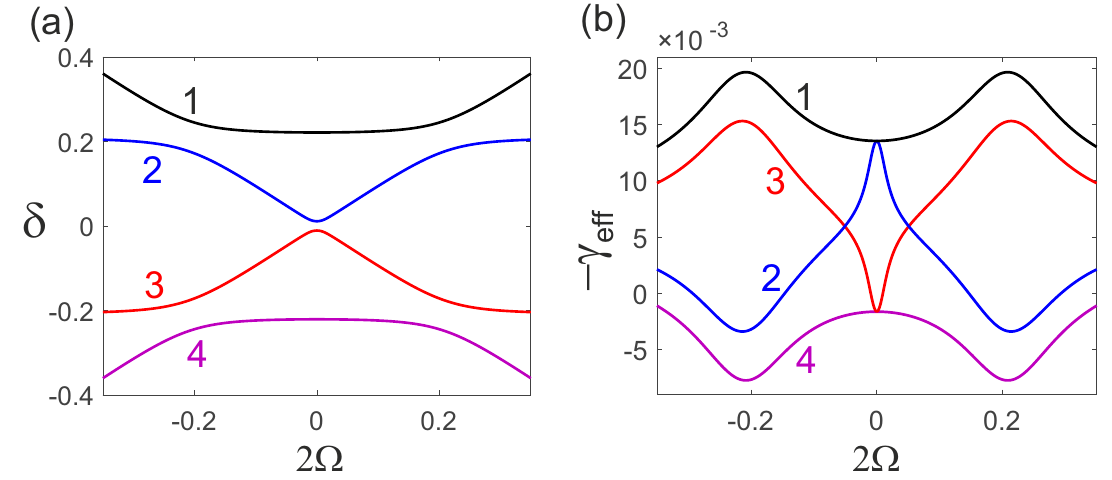}
\caption{The real (a) and imaginary (b) parts of the frequencies of the polariton eigenmodes as functions of the angular velocity of the potential~$\Omega$. 
\label{FIG_eigfreq_Om}}
\end{figure}

For further analysis, we rely on the numerically calculated dependencies of the eigenfrequencies on the angular velocity of the rotating potential. These dependencies are shown in  Fig.~\ref{FIG_eigfreq_Om}. It is evident that there exist four modes characterized by distinct effective decay rates~$\gamma_{\text{eff}}$. These differing decay rates emerge due to the dissipative contribution introduced by the rotating potential, characterized by the coupling constant~$\eta$.
The effective decay rate $\gamma_{\text{eff}}$ encompasses a range of non-conservative processes that influence the mode. These processes include not only losses of polaritons but also the filling of the mode due to the presence of an external pump.
Due to the pump, the dissipation rate $\gamma_{\text{eff}}$ can take negative values, which correspond to the temporal growth of the mode. From the physical point of view, this growth signifies the condensation of the incoherent reservoir excitons into the coherent polariton mode. It is worth mentioning that in our model, the mode with the highest frequency experiences the fastest growth.
Given our focus on this mode in further analysis, we extract its associated eigenvector $\vec X=\left( X_{\uparrow +}, X_{\uparrow -}, X_{\downarrow +}, X_{\downarrow -}\right)^{\text{T}}$ to emphasize its characteristics

The structure of the stationary polariton state is expected to resemble that of the fastest growing mode when the latter effectively suppresses other modes, and the stationary condensate density remains low enough to alter the field distributions in the polariton state significantly.
To validate the similarity between the structures of the fastest growing mode and the stationary polariton state, we conducted a comparison of their normalized angular momenta $\mathcal{M} _{\uparrow, \downarrow, \text{tot}}$ and population difference in the orthogonal circularly polarized components~$\Delta N /N_{\text{tot}}$, as depicted in Figs.~\ref{FIG_AM_ndiff_dpendns}(c,d). The indicated quantities for the fastest growing mode were calculated from~Eq.~\eqref{EqAngMandNviaAmplC} after substituting~$C$~with~$X$.
The dependencies of these quantities on the angular velocity~$\Omega$ are shown in Fig.~\ref{FIG_AM_ndiff_dpendns}(c,d).
It is evident that both the angular momenta and the population differences of the stationary nonlinear states are quite close to those of the fastest growing linear mode. 
This robust similarity strongly suggests that the former inherits the structure of the latter.

\section{Perturbative treatment of the nonlinear problem \label{Nonlinear_perturbation}}

We now extend our investigation by considering the evolution of polaritons within the coupled modes approximation while introducing nonlinear effects through the framework of perturbation theory.
We amend Eq.~\eqref{EqEigEq} by incorporating the nonlinear term~$\vec N (\vec C)$ from the right hand side:
\begin{eqnarray}
\frac{d}{dt} \vec C = i \hat L \vec C + \vec N (\vec C), \label{pt_theorNL}
\end{eqnarray}
The nonlinear term can be derived by the calculation of the nonlinear term and then by the projection on the modes of the axially symmetric conservative problem. After a simple algebra one can obtain
\begin{widetext}
\begin{eqnarray}
\vec N (\vec C)= \begin{pmatrix}
(i\alpha - \alpha_d)(|C_{\uparrow+}|^2+2|C_{\uparrow-}|^2) C_{\uparrow+}  
+i\alpha_x [ ( |C_{\downarrow+}|^2+|C_{\downarrow-}|^2 )C_{\uparrow+} + C_{\uparrow-}C_{\downarrow+}C_{\downarrow-}^{*} ]  \\
(i\alpha -\alpha_d)(|C_{\uparrow-}|^2+2|C_{\uparrow+}|^2)C_{2} 
+i\alpha_x[ ( |C_{\downarrow+}|^2+|C_{\downarrow-}|^2 )C_{\uparrow-} + C_{\uparrow+}C_{\downarrow-}C_{\downarrow+}^{*} ] \\
(i\alpha -\alpha_d)(|C_{\downarrow+}|^2+2|C_{\downarrow-}|^2)C_{\downarrow+}
 +i\alpha_x[ ( |C_{\uparrow+}|^2+|C_{\uparrow-}|^2 )C_{\downarrow+} + C_{\downarrow-}C_{\uparrow+}C_{\uparrow-}^{*} ] \\
  (i\alpha -\alpha_d(|C_{\downarrow-}|^2+2|C_{\downarrow+}|^2)C_{\downarrow-}
+i\alpha_x[ ( |C_{\uparrow+}|^2+|C_{\uparrow-}|^2 )C_{\downarrow-} + C_{\downarrow+}C_{\uparrow-}C_{\uparrow+}^{*} ]
\end{pmatrix}.
\end{eqnarray}
\end{widetext}

As discussed in the main part of the paper, the rotating potential couples the modes. To get analytical results on the nonlinear dynamics, it is convenient to write the equation analogous to   \ref{pt_theorNL} but for the amplitudes of the super-modes --- the eigenmodes of the linear problem accounting for all real potentials. The dissipative and nonlinear effects will be treated perturbatively in terms of the amplitudes of the super-modes. Thus, in our treatment, we assume that the linear conservative terms included in the real part of the operator $\hat {L}$, namely the TE-TM splitting ($\sigma$) and the conservative mode coupling originating from the rotating potential ($\eta _r$), significantly outweigh the increments (decrements) of the modes as well as the contributions from all nonlinear terms. 

Then we split the right-hand side of Eq.~\eqref{pt_theorNL} into two components: the conservative linear part, which provides the fundamental solutions to a linear problem, and the supplementary part, encompassing dissipative and nonlinear effects that we treat as perturbations.
Then, in the leading approximation order, the linear part is characterized by the operator
\begin{equation}
\label{EqL0}
\hat L_0=
\left( \begin{array}{cccc}
0 & \eta _r & 0 & \sigma \\
\eta _r & 2\Omega & 0 & 0 \\
0 & 0 & -2\Omega & \eta _r \\
\sigma & 0 & \eta _r & 0
\end{array} \right)
\end{equation}
including the TE-TM splitting ($\sigma$) and conservative coupling due to the rotating potential ($\eta _r$).
The non-conservative effects are of the next perturbation order and characterized by the operator
\begin{equation}
\label{EqL1}
\hat L_1=
i \left( \begin{array}{cccc}
\gamma_{0} & \eta _i & 0 & 0 \\
\eta _i & \gamma_{0} & 0 & 0 \\
0 & 0 & \gamma_{0} & \eta _i \\
0 & 0 & \eta _i & \gamma_{0}
\end{array} \right)
\end{equation}
including the losses of the modes ($\gamma _0$) and gain from the rotation-induced mode mixing~($\eta _i$).
In~\eqref{EqL0} and~\eqref{EqL1},  $\eta _r = \text{Re}(\eta)$ and $\eta _i = \text{Im}(\eta)$.

We look for a solution in the following form:
\begin{equation}
\label{EqProjBasis}
\vec C = \sum_{k=1}^{4} a_k \vec Y_k e^{i \omega_k t},
\end{equation}
where $\omega_k$ are the eigenvalues of $\hat L_0$. One should underline that $\omega$ and $\vec Y$ are purely real. For the sake of convenience, we normalize $\vec Y$ such that~$|\vec Y|^2=1$. 
We assume that the mode $\vec Y_1$ is the fastest growing and thus dominating mode. 

Then we project Eq.~\eqref{pt_theorNL} onto the basis of eigenvectors $\vec Y$ of the operator $\hat{L} _0 = \text{Re}(\hat L)$ and consider the terms $\hat{L} _1 \vec{C} = \text{Im} (\hat{L}) \vec{C}$ and $\vec{N} (\vec{C})$ as perturbations. For our purposes, it is sufficient to derive the equation for the amplitude of the fastest growing super-mode, assuming that the other super-modes are negligibly small. 
The equation reads:
\begin{eqnarray}
\dot a_1=\left( \vec Y_1 ^{\text{T}} \hat L_1 \vec Y_1\right)a_1 + \left( \vec Y_1 ^{\text{T}} \vec N_{y_1}\right)|a_1|^2a_1,
\end{eqnarray}
where $\vec N_{y_1}$ is $\vec N$ calculated for $\vec Y_1$.

From this equation, it is easy to find the stationary amplitude of the mode:
\begin{multline}
|a_{1 \text{st}}|^2 = \text{Re} \left( \frac{  \vec Y_1 ^{\text{T}} \hat L_1 \vec Y_1 }{ \vec Y_1 ^{\text{T}} \vec N_{y_1}} \right)\\
= - \frac{2 \eta _i (Y_{1\uparrow +}Y_{1\uparrow -}+Y_{1\downarrow +}Y_{1 \downarrow -})+\gamma_{\text{eff}} }{\alpha_d\left( Y_{1 \uparrow +}^4+Y_{1 \uparrow -}^4+Y_{1 \downarrow +}^4+Y_{1 \downarrow -}^4\right)},
\label{am_stationary}
\end{multline}
where $\vec Y_{1}=\left( Y_{1\uparrow +}, Y_{1\uparrow -}, Y_{1\downarrow +}, Y_{1\downarrow -}\right)^T$.
The correction to the frequency of the state is found as follows:
\begin{equation}
\delta_{\text{nl}}=\frac{\alpha}{\alpha_d} \left[ 2 \eta _i (Y_{1\uparrow +}Y_{1\uparrow -}+Y_{1\downarrow +}Y_{1\downarrow -})+\gamma_{\text{eff}}\right].
\end{equation}

\begin{figure}[tb!]
\centering\includegraphics[width=\columnwidth]{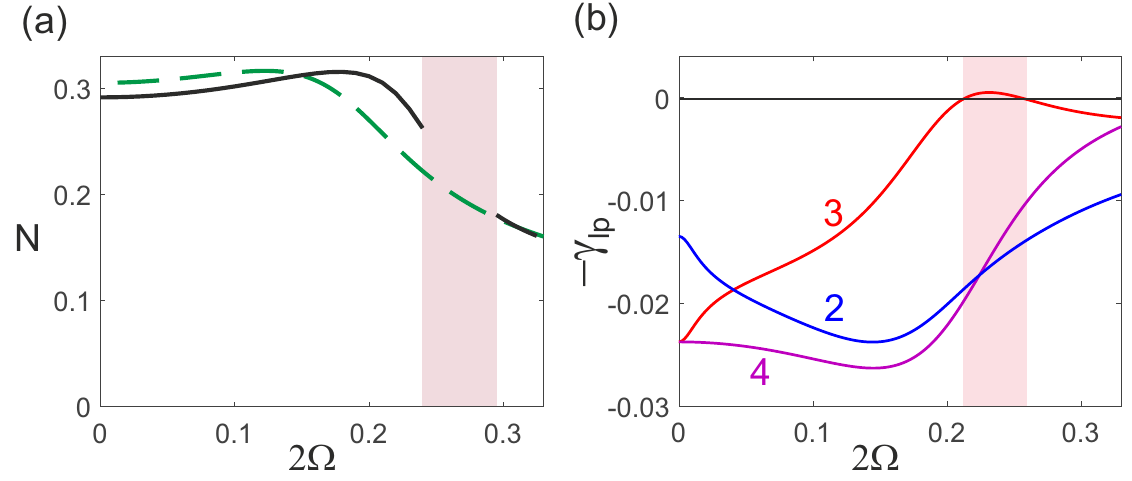}
\caption{(a) The dependencies of the polariton number in the stationary states, calculated within the coupled modes approximation, on the rotation velocity $\Omega$ of the potential. The black solid line corresponds to the direct numerical simulations of Eq.~\eqref{EqCoupled_modeC}, while the dashed green line is for the results of the perturbation theory treatment. The shaded rectangle marks the range of $\Omega$ where single-frequency solutions of Eqs.~\eqref{EqCoupled_modeC} do not exist.
(b) The effective increments of the linear excitations on the backgroung of the stationary states calculated through the perturbation theory. The colors of the lines correspond to those in~ Fig.~\ref{FIG_eigfreq_Om}. 
The shaded rectangle shows the range of $\Omega$ where one of the modes (red) has a positive increment. 
\label{FIG_pertutb_N_Om}}
\end{figure}

The total number of polaritons is expressed through the amplitudes~$a_k$ as~${N = \sum _k |a_k|^2}$.
The perturbatively found dependency~$N$ on the angular velocity of the potential $\Omega$ is illustrated in Fig.~\ref{FIG_pertutb_N_Om}(a) with a green dashed line. 
Clearly, perturbation theory provides a satisfactory approximation for the parameters employed in direct numerical simulations. We have verified that this agreement becomes better for lower values of the effective gain $\gamma_{\text{eff}}$.

\section{The stability of a single-mode solution and oscillating states \label{oscillating behavior}}

The previously found single-mode solution may be unstable, with other modes potentially growing and significantly affecting the solution. To assess its stability, we linearize~\eqref{pt_theorNL} and derive an equation for small excitation $\vec{\xi}$ on the background of the stationary solution $a_{1 \text{st}} \vec{Y_{1}}$. 
Excitations with eigenfrequencies substantially detuned from the frequency of the stationary solution, when the detuning is much greater than the characteristic timescales of the dissipative and nonlinear terms, can be described by the following equation:  
\begin{eqnarray}
\frac{d \vec \xi}{dt}  = i \hat L_0 \vec \xi+\hat M  \vec \xi. \label{pt_theor_sm_exct}
\end{eqnarray}
Here $\hat{M} = \hat{L}_1  +|a_{1 \text{st}}|^2 ( \partial \vec{N} / \partial \vec{C} )$, where $\partial \vec{N} / \partial \vec{C}$ represents the Jacobian evaluated at ${\vec{C} = \vec{Y} _1}$. It is important to note that this equation does not account for parametric effects. 

Considering $\hat{M}$ as a perturbation, we can search for a solution in the form $\vec{\xi} _k= \vec{Y}_k e^{i\omega_k t}e^{i\delta_k t} $ and derive expressions for~$\delta_k$:
\begin{eqnarray}
\delta_k= -i \left( \vec Y_k ^{\text{T}} \hat M \vec Y_1 \right). 
 \label{pt_theor_increment}
\end{eqnarray}
One should note that \eqref{pt_theor_increment} is not applicable to  excitaions having the same structure as the fastest growing mode~$\vec{Y} _1$, so this formula is valid only for~$k \neq 1$.

The perturbatively obtained dependencies of the increments of the modes $\gamma_{\text{lp} \, k} = \text{Im} (\delta_{k})$ are shown in Fig.~\ref{FIG_pertutb_N_Om}(b).
The negative value of $\gamma_{\text{lp}\, k}$ indicates that the corresponding mode $k$ grows in time. 
One can see that for the mode labeled as ``3'' in the figure (red curve), there is a range of angular velocities~$\Omega$, where it exhibits a positive increment. {This implies that the stationary state cannot be defined by the fastest growing linear mode alone, and one can anticipate the oscillatory dynamics.} 

Note that the range of $\Omega$ for the oscillating behavior is shifted compared to what was observed in direct numerical simulations. This discrepancy can be attributed to the relatively large magnitude of the perturbation for the parameters used in direct numerical simulations. We have verified that most of this discrepancy arises from the nonlinear shift of the modes' eigenfrequencies, and that the agreement improves as pump intensity decreases.  

Now, let's consider whether the oscillatory state can be considered as coexistence of modes with the structures of the fastest growing linear mode and the mode that is not suppressed by the former. According to the perturbation theory, the second mode constituting the oscillating state should be similar to the linear mode labeled as~``3'' in~Fig.~\ref{FIG_eigfreq_Om}.

\begin{figure}[tb!]
\centering\includegraphics[width=0.9\columnwidth]{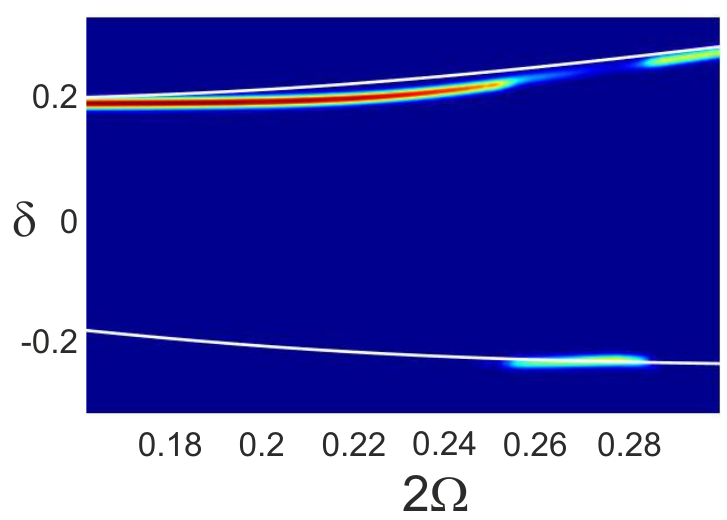}
\caption{The temporal spectrum of the stationary state as a function of the angular velocity of the potential~$\Omega$. The simulations were conducted within the coupled mode approximation. The white curves show the red-shifted dependencies of the real parts of the eigenfrequencies of the two linear modes, including the fastest growing mode shown by the black curve and the state indicated by the red curve in Fig.~\ref{FIG_eigfreq_Om}(a). The eigenfrequencies are shifted downward by~$0.04$. 
\label{FIG_the_spectrum}}
\end{figure}

Through direct numerical simulations, we have obtained the temporal spectra of the stationary states for various angular velocities of the potential $\Omega$, as depicted in Fig.~\ref{FIG_the_spectrum}. In this plot, an interval of $\Omega$ exists where the system exhibits two-frequency dynamics, when some polaritons oscillate at a higher frequency $\delta_1$, while others oscillate at a lower frequency~$\delta_2$. The white curves on this spectrum correspond to the eigenfrequencies of the modes ``1'' (fastest growing) and ``3'' shifted down by~$0.04$. This shift is approximately equal to the shift observed in these modes due to the nonlinearity effect. Notably, these spectral lines closely follow the dependencies of the linear eigenfrequencies on the potential rotation velocity.

\begin{figure}[tb!]
\centering\includegraphics[width=\columnwidth]{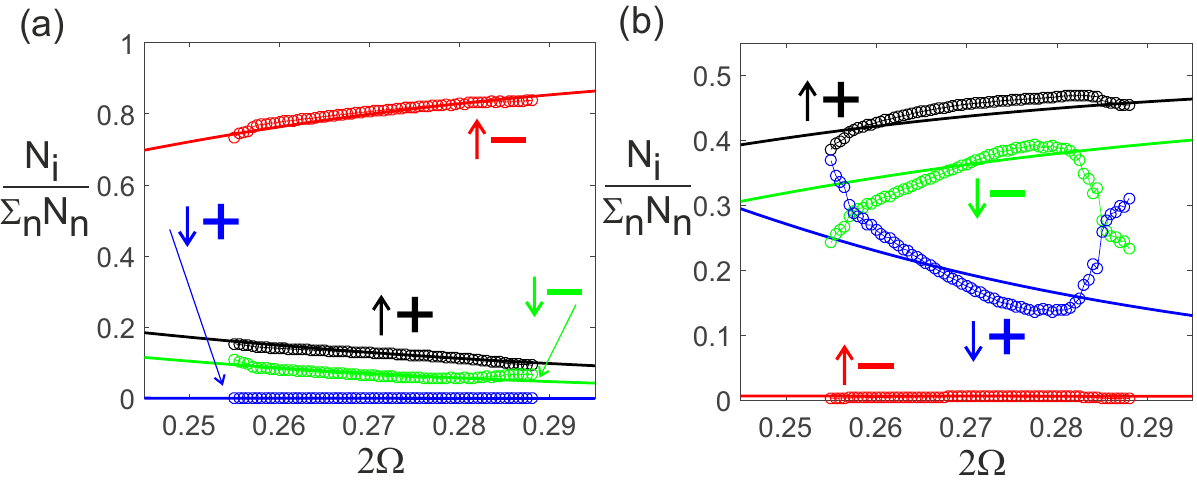}
\caption{The structure of the components corresponding to the upper (a) and the lower (b) spectral lines shown in Fig.~\ref{FIG_the_spectrum}. The open circles correspond to the data extracted from the numerical simulations within the coupled modes model, while the solid lines correspond to the quantities calculated for the eigenmodes shown by black (for panel (a) ) and red (for panel (b) ) colors in  Fig.~\ref{FIG_eigfreq_Om}. The field structure is characterized by the rations  ${N_i \left/ \sum_n N_n \right.} = {\left. |C_i|^2 \right/ |\vec C|^2}$ shown as functions~of~$\Omega$. 
\label{FIG_the_oscill_modes}}
\end{figure}

In our numerical experiments, we have the flexibility to isolate the first or second spectral lines, enabling us to analyze the polarization and angular momentum of polariton states associated with each of the frequencies. In essence, we represent the stationary solution as $\vec C = c_1 \vec Z_{1} e^{i\delta_1 t}+ c_2 \vec Z_1 e^{i\delta_2 t} $. Here, $c_{1, 2}$ denote the amplitudes determining the number of polaritons, while the vectors $\vec Z_{1,  2}$, normalized such that ${|\vec Z_{1,2}| = 1}$, describe how polaritons with frequencies $\delta_{1,2}$ are distributed among the states of $\uparrow$ and $\downarrow$ polarizations, as well as among positive ($+$) and negative ($-$) angular momenta.
The dependencies of the reduced occupations of the states $|Z_{ij}|^2 / |\vec{Z}|^2$, where $i=\uparrow, \downarrow$ and $j = \pm$, are shown in Fig.~\ref{FIG_the_oscill_modes} as functions of $\Omega$ for polaritons of the frequency~$\delta_1$ (a) and~$\delta_2$~(b).
One can see that, indeed, the distribution of polaritons with frequency $\delta_1$ across polarizations and angular indices closely resembles that of the fastest growing linear mode, which has the highest eigenfrequency. At the same time, the dependencies plotted in panel (b) confirm that the polaritons with frequency $\delta_2$ exhibit an order parameter similar to that of the mode that is not suppressed by the fastest growing mode, which is labeled as~``3."

Finally, let us provide some estimations to clarify the previously discussed effect, which is the limited influence of polariton transfer between polarizations on the total number of polaritons. The polarization oscillations arise due to the interaction between two coexisting modes, and therefore the period of the polarization oscillations $T_{\text{osc}}$ is  inversely proportional to the frequency difference between these coexisting modes. For the modes involved, this frequency difference is primarily defined by the TE-TM splitting and can be estimated as $\Delta \delta \approx 2\sigma$.  On the other hand, the characteristic timescale for the evolution of the polariton number $T_{\text{dns}}$ is inversely proportional to  $|\gamma_0 + \eta _i|$. This allows us to estimate the ratio of these characteristic times as $\left. T_{\text{dns}} \right/  T_{\text{osc}} \approx \left.2|\sigma| \right/ |\gamma_0+\eta_i|$. For the used parameters, this ratio is approximately 20, which explains why our numerical simulations do not show significant oscillations in the total polariton number despite the inter-polarization polariton transfer.

\bibliography{main_text_yulin}
\end{document}